\documentclass{iau379}

\usepackage{graphicx}
\usepackage{amsmath}
\usepackage{multirow}

\begin{document}

\newcommand{\kms}{\ensuremath{\mathrm{km}\,\mathrm{s}^{-1}}}
\newcommand{\galunits}{\ensuremath{\mathrm{km}\,\mathrm{s}^{-1}\,\mathrm{kpc}^{-1}}}
\newcommand{\galacc}{\ensuremath{\mathrm{km}^2\,\mathrm{s}^{-2}\,\mathrm{kpc}^{-1}}}
\newcommand{\MLsun}{\ensuremath{\mathrm{M}_{\odot}/\mathrm{L}_{\odot}}}
\newcommand{\Lsun}{\ensuremath{\mathrm{L}_{\odot}}}
\newcommand{\Msun}{\ensuremath{\mathrm{M}_{\odot}}}
\newcommand{\Aunits}{\ensuremath{\mathrm{M}_{\odot}\,\mathrm{km}^{-4}\,\mathrm{s}^{4}}}
\newcommand{\surfdens}{\ensuremath{\mathrm{M}_{\odot}\,\mathrm{pc}^{-2}}}
\newcommand{\voldens}{\ensuremath{\mathrm{M}_{\odot}\,\mathrm{pc}^{-3}}}
\newcommand{\gevcc}{\ensuremath{\mathrm{GeV}\,\mathrm{cm}^{-3}}}
\newcommand{\LCDM}{$\Lambda$CDM}
\newcommand{\ML}{\ensuremath{\Upsilon_*}}
\newcommand{\Mst}{\ensuremath{M_*}}
\newcommand{\Mg}{\ensuremath{M_g}}
\newcommand{\Mb}{\ensuremath{M_b}}
\newcommand{\Mhalo}{\ensuremath{M_{\mathrm{200}}}}
\newcommand{\Vhalo}{\ensuremath{V_{\mathrm{200}}}}
\newcommand{\Vf}{\ensuremath{V_f}}
\newcommand{\gobs}{\ensuremath{\mathrm{g}_{\mathrm{obs}}}}
\newcommand{\gtot}{\ensuremath{\mathrm{g}_{\mathrm{tot}}}}
\newcommand{\gbar}{\ensuremath{\mathrm{g}_{\mathrm{bar}}}}
\newcommand{\azero}{\ensuremath{\mathrm{a}_{0}}}
\newcommand{\gdagger}{\ensuremath{\mathrm{g}_{\dagger}}}
\newcommand{\Hunits}{\ensuremath{\mathrm{km}\,\mathrm{s}^{-1}\,\mathrm{Mpc}^{-1}}}
\newcommand{\apj}{Astrophys.\ J.}
\newcommand{\apjl}{Astrophys.\ J.}
\newcommand{\aj}{Astron.\ J.}
\newcommand{\mnras}{Mon.\ Not.\ Royal Astron.\ Soc.}
\newcommand{\aap}{A \& A}
\newcommand{\aapr}{Astron.\ \& Astrophys.\ Review}
\newcommand{\araa}{Ann.\ Rev.\ Astron.\ \& Astrophys.}
\newcommand{\pasa}{Pub.\ Astron.\ Soc.\ Australia}
\newcommand{\jcap}{J.\ Cosmology \& Astroparticle Phys.}
\newcommand{\prl}{Phys.\ Rev.\ Lett.}

\lefttitle{McGaugh}
\righttitle{The Local Group as seen from the outside looking in}

\jnlPage{1}{7}
\jnlDoiYr{2023}
\doival{10.1017/xxxxx}

\aopheadtitle{Proceedings of IAU Symposium 379}
\editors{P. Bonifacio,  M.-R. Cioni, F. Hammer, M. Pawlowski, and S. Taibi, eds.}

\title{Local Group Galaxies from an External Perspective}

\author{McGaugh S.S.$^1$}
\affiliation{$^1$ Case Western Reserve University, 10900 Euclid Ave., Cleveland, OH, USA}

\begin{abstract}
I discuss Local Group galaxies from the perspective of external galaxies that define benchmark scaling relations. Making use of this information leads to a model for the Milky Way that includes bumps and wiggles due to spiral arms. This model reconciles the terminal velocities observed in the interstellar medium with the rotation curve derived from stars, correctly predicts the gradual decline of the outer rotation curve, and extrapolates well out to 50 kpc. Rotationally supported Local Group galaxies are in excellent agreement with the baryonic Tully-Fisher relation. Pressure supported dwarfs that are the most likely to be in dynamical equilibrium also align with this relation. Local Group galaxies thus appear to be normal members of the low redshift galaxy population. There is, however, a serious tension between the dynamical masses of the Milky Way and M31 and those expected from the stellar mass–halo mass relation of abundance matching.
\end{abstract}

\begin{keywords}
Local Group (929), Milky Way Galaxy (1054), Dwarf galaxies (416), Galaxy masses (607)
\end{keywords}

\maketitle

\section{Introduction}

Local Group galaxies are accessible to detailed study with large samples of individual stars.
This information is complemented by information from external galaxies, which obey strong kinematic scaling relations \citep{2020IAU,L2022}. 
These include the Baryonic Tully-Fisher Relation \citep[BTFR:][]{JSH0}, the Central Density Relation \citep{L2016}, 
and the Radial Acceleration Relation \citep[RAR:][]{RAR,L2017}.
Local Group galaxies prove to be completely normal members of the general galaxy population by these standards.

\section{The Milky Way}

The Milky Way is of obvious interest as our home galaxy, yet our place within it is a source of constant frustration: we cannot see it from `above' as we do external spirals.
We nevertheless infer that it is a barred spiral thanks to decades of work employing a broad spectrum of methods \citep{BG2016}.
In external galaxies, spiral features cause bumps and wiggles in the light profile that correspond to
features in the rotation curve \citep{renzo}. This correspondence is quantified by the RAR.

\begin{figure}[t]
  \centerline{\vbox to 6pc{\hbox to 10pc{}}}
  \includegraphics[scale=0.27]{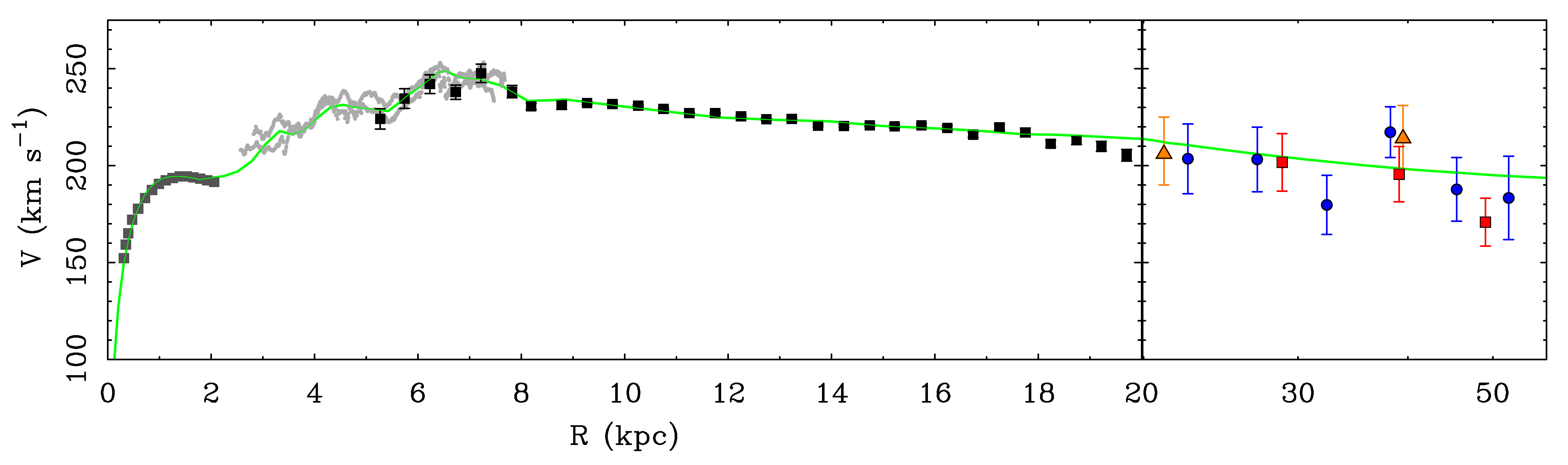}
  \caption{The rotation curve of the Milky Way. Light gray points are the terminal velocity data \citep{MG2007,MG2016} to which the model was fit \cite[green line]{M2018}.
  The model was not fit to the data of \cite{MGW2017} (dark gray squares at $R < 2.2$ kpc) or \cite{E2019} (black squares), but is nicely consistent with them \citep[][]{M2019}. 
  It also extrapolates well out to 50 kpc as traced by  
  K giants (red squares) and BHB stars (blue circles) \citep{Bird2022} and by globular clusters \citep[orange triangles]{Watkins2019}.}
    \label{fig:MWRC}
\end{figure}

The RAR can be applied to the Milky Way to build a mass model for the stellar disk \citep{M2016}. 
The terminal velocities observed by \cite{MG2007} and \cite{MG2016} were fit in detail between $3 < R < 8$ kpc. 
The resulting model is shown in Fig.\ \ref{fig:MWRC}, updated to the distance scale \citep{M2018} indicated by the \cite{GRAVITY}.

\begin{figure}[t]
  \centerline{\vbox to 6pc{\hbox to 10pc{}}}
  \includegraphics[scale=0.22]{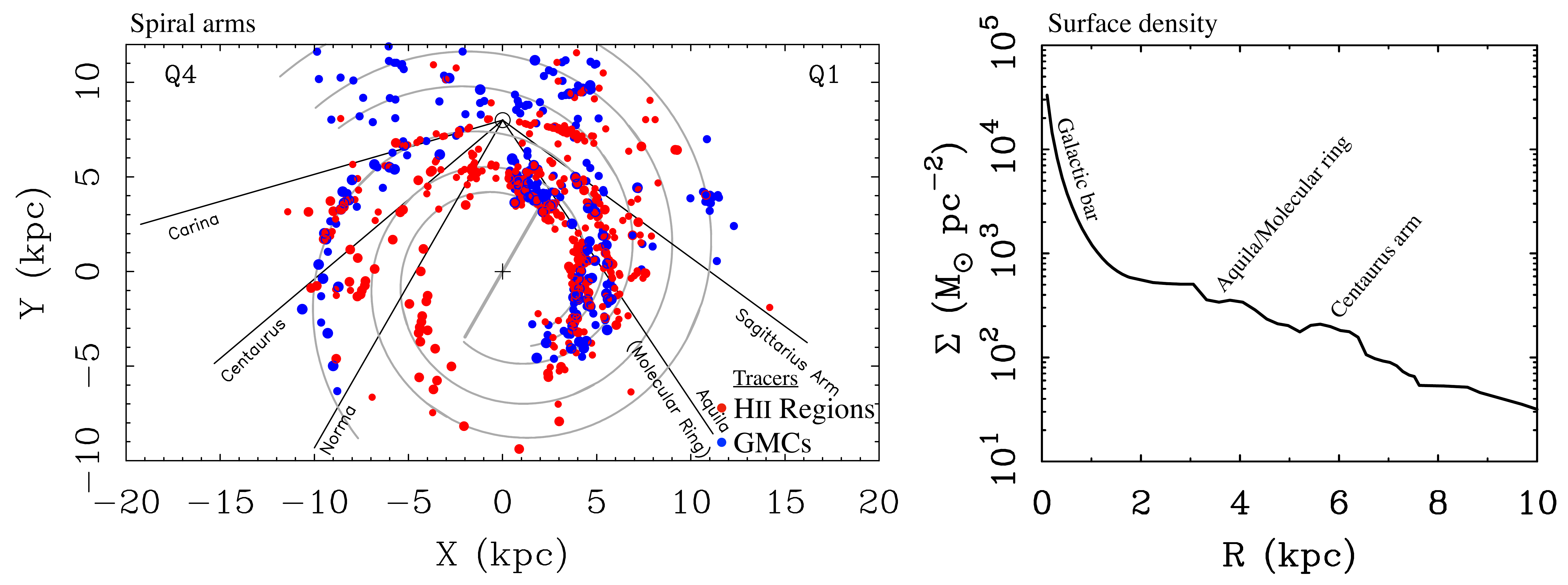}
  \caption{Spiral structure (left) and surface density profile (right) of the Milky Way. The bumps and wiggles inferred kinematically correspond well to the features known from
  tracers like HII regions (red points) and Giant Molecular Clouds (blue points) \citep{M2016}.}
    \label{fig:MWspiral}
\end{figure}

Fig.\ \ref{fig:MWspiral} shows the surface density corresponding to the rotation curve seen in Fig.\ \ref{fig:MWRC}.
The model has bumps and wiggles in the profile that are reminiscent of those seen in external spirals. Indeed,
these kinematically inferred features correspond well to spiral arms that are known from traditional tracers (Fig.\ \ref{fig:MWspiral}).

The model \citep{M2018} correctly predicted the slope $dV/dR = -1.7\;\galunits$ of the outer rotation curve subsequently found by \cite{E2019}.
It extrapolates well to data for globular clusters \citep{Watkins2019} and halo stars \citep{Bird2022} at still larger radii,
and is nicely consistent with the inner mass distribution inferred by \cite{MGW2017} without that region having been specifically fit.
This is a testament to how tightly the mass distribution is coupled to that of the light: the latter is predictive of the former \citep{2020Galax}. 

\begin{table}[h]
 \centering
 \caption{Mass Estimates}\label{tab:mass}
 {\tablefont\begin{tabular}{@{\extracolsep{\fill}}llccccl}
    \midrule	
    Object & Model & \Mst & \Mhalo & $c$ & \Vhalo & Ref. \\
     & & $10^{10}\;\Msun$ & $10^{10}\;\Msun$ & & \kms & \\
    \midrule
  Milky Way   & NFW from RAR & \phantom{1}6.2 & 139 & \phantom{1}8.1 & 162 & \cite{M2018} \\ 
    M31 & NFW+SSP & 13.5 & 160 & 10.1 & 169 & \cite{Chemin2009} \\
    M31+MW & Straight sum & 19.7 & 299 & \dots & \dots & \cite{MvD2021} \\
    \midrule
    \end{tabular}}
\end{table}

The stellar rotation curve derived by \cite{E2019} is formally discrepant from the terminal velocities \citep{MG2007} at high significance.
The model reconciles the two through numerical evaluation of the tracer term $\partial \ln \nu/\partial \ln r$ in the Jeans equation \citep{M2019}.
The apparent discrepancy is an artifact of the usual assumption of an exponential disk, which does not account for the bumps and wiggles in $\nu(r)$.
Galactic data have reached a level of precision that these can no longer be neglected. 

We can query the model for the halo mass it implies (Table \ref{tab:mass}). 
To be specific, I quote $M_{200}$ for the standardization it provides, and the estimate I obtain
is consistent with the range found by others \citep{Wang2020}. 
However, we should bear in mind that this mass is a rather notional quantity that involves 
an extrapolation to a radius that is not directly constrained by data using a model \citep{NFW} that 
persistently provides a poor depiction of observational reality. 

A more robust quantity is the enclosed mass $M(R < 50\;\mathrm{kpc}) = 44 \times 10^{10}\;\Msun$.
This agrees well with halo star \citep{Bird2022} and globular cluster \citep{Watkins2019} data.
Going farther out is complicated by the perturbing influence of the LMC.
Indeed, if we take seriously the downturn in the Gaia-derived rotation curve 
at $R \approx 19\;\mathrm{kpc}$ \citep{Wang2023}, a much lower mass is obtained \citep{SL2023}.
It is difficult to envision a model that is consistent with both this downturn and the data for halo stars.

Even in the absence of these considerations, the total mass of the Milky Way remains a fraught issue. 
Weak lensing observations of external galaxies \citep{Brouwer} indicate that rotation curves persist in being approximately flat 
to hundreds of kpc, reaching beyond the expected virial radii of bright spirals. 
This is not consistent with an NFW profile, but it is consistent with the RAR, which is what works well above. These are not the same.

\section{The Local Group}

The Milky Way falls on the BTFR defined by external galaxies (Fig.\ \ref{fig:LGBTF}). Indeed, all of the rotationally supported Local Group galaxies do so,
within the uncertainties: there is no indication that they are abnormal in this regard. This provides both a constraint --- nothing 
particularly peculiar is permitted --- and an opportunity: if we can calibrate the dynamical masses of Local Group galaxies, we can infer those of more distant galaxies. 

\subsection{Dwarf Spheroidal Galaxies and Tully-Fisher}

\begin{figure}[t]
  \centerline{\vbox to 6pc{\hbox to 10pc{}}}
  \includegraphics[scale=0.6]{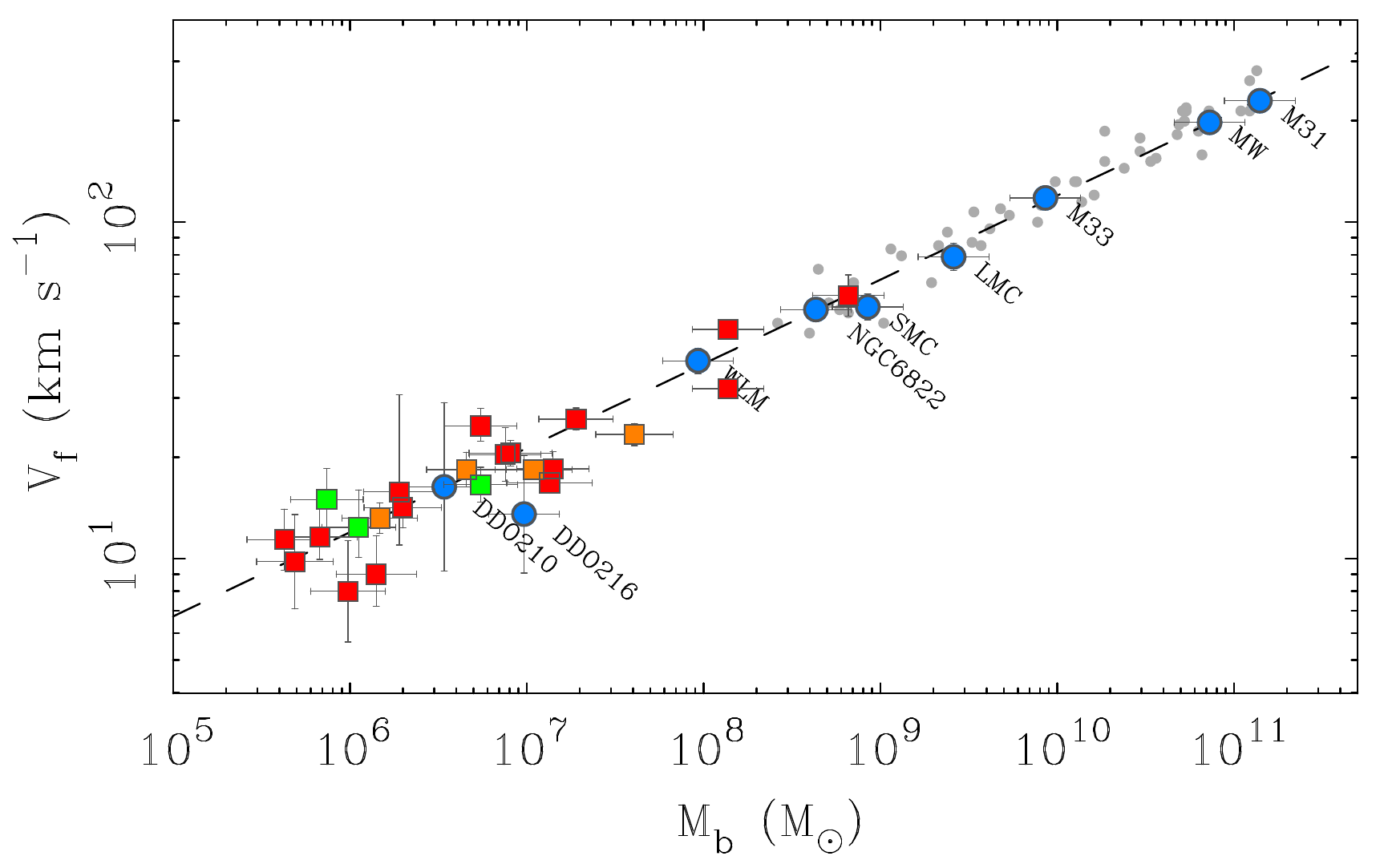}
  \caption{The baryonic Tully–Fisher relation of Local Group galaxies (blue circles) follows that defined by external galaxies \cite[grey points][]{JSH0}.
  This extrapolates to pressure supported dwarfs (squares) when $\Vf = 2 \sigma_*$ \citep{LGTF}.
  Red squares are satellites of M31, orange squares are satellites of the Milky Way; green squares are unaffiliated \citep{McC2021}.}
  \label{fig:LGBTF}
\end{figure}

The BTFR may further serve as a method to unite the mass scales of pressure supported early type dwarf galaxies with their rotationally supported late type brethren.
Typically we measure only a single, bulk velocity dispersion $\sigma_*$ for dwarf spheroidals. 
The dwarf spheroidals of the Local Group fall into two populations in the BTFR plane \citep{MW2010}:
the $L$-$\sigma_*$ data for classical dwarfs parallels the BTFR, while for the ultrafaint dwarfs $\sigma_*$ is uncorrelated with $L$, seeming to hit a floor.
As a consequence of their relatively high velocity dispersions, the latter appear to be very dense and hence safe from tidal influences \citep{Pace2022}. 
This may be an illusion of the assumption of dynamical equilibrium, as it seems unlikely that these tiny dwarfs can survive 
unmolested deep in the potential well of the Milky Way while the Milky Way itself is being strongly perturbed by the LMC. 
Excepting the (primarily ultrafaint) dwarfs that may be subject to tidal influence by the criterion of \cite{Bellazzini1996},
the remaining dwarfs of the Local Group fall nicely on the BTFR defined by rotating galaxies given a simple transformation
$\Mst = \ML L_V$ and $\Vf = \beta_c \sigma_*$ with $\log \beta_c = 0.25 \log\ML +0.226$ \citep{LGTF}.
This works out to $\beta_c =2$ for $\ML = 2\;\MLsun$.
This quantifies the circular velocity that is equivalent to the line of sight velocity dispersion in an average sense.
The strong regularity seen in Fig.\ \ref{fig:LGBTF} is surely teaching us something about the dynamical masses of galaxies.

\subsection{Kinematic Masses and Abundance Matching}

A common method for relating the observed stellar masses of galaxies with their host dark matter halos is 
Abundance Matching \citep[AM:][]{Behroozi2013,Moster2013,Kravtsov2018,Mowla2019}.
Indeed, this has become an essential ingredient for understanding galaxies in \LCDM.
The Local Group provides an ideal opportunity to test these relations with independent kinematic data.

\begin{figure}[t]
  \centerline{\vbox to 6pc{\hbox to 10pc{}}}
    \includegraphics[scale=.3]{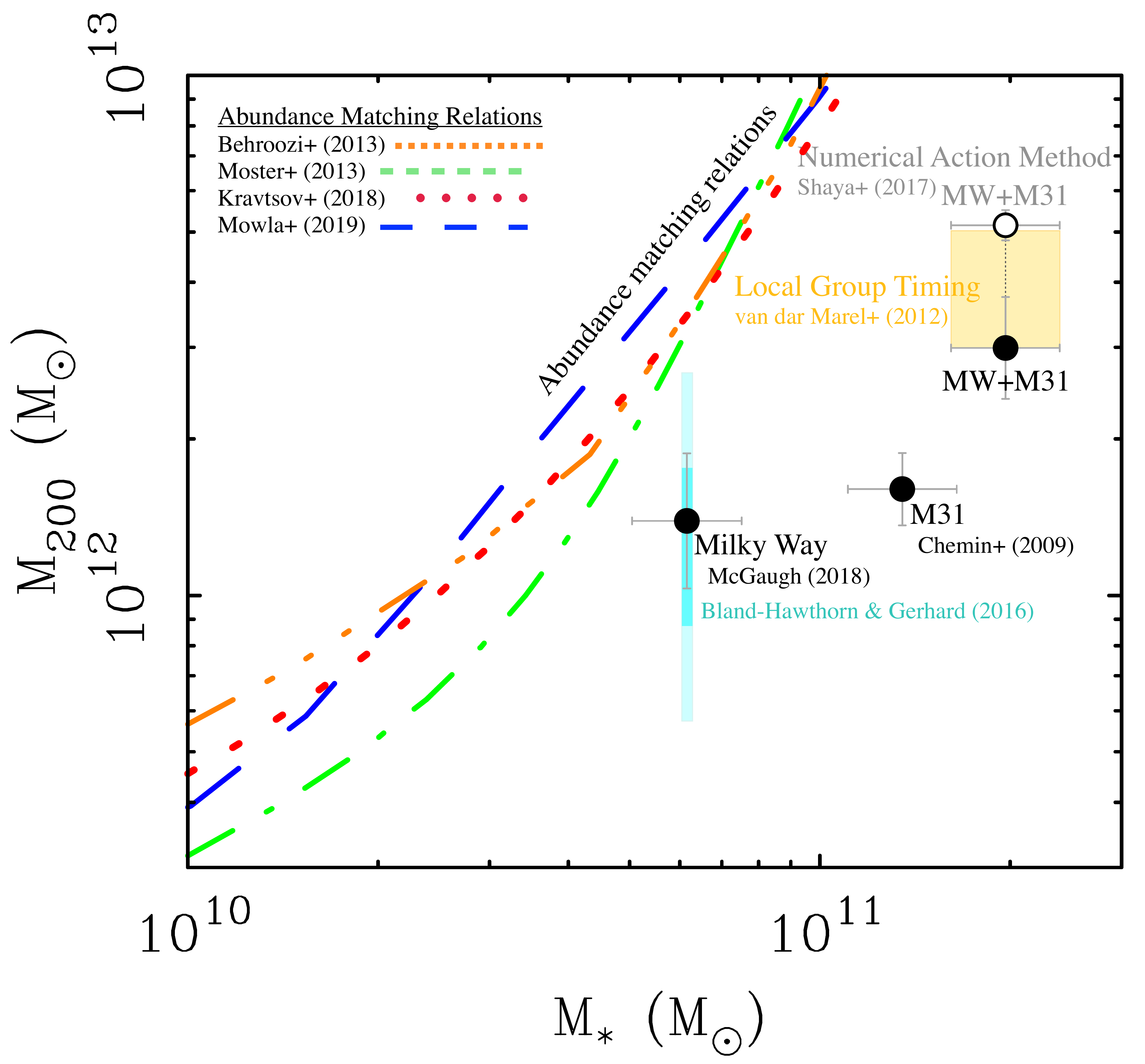}
  \caption{Stellar mass--halo mass relations in the context of the Local Group \citep{MvD2021}. Filled points are from rotation curve fits to the Milky Way \citep{M2018}, M31 \citep{Chemin2009}, and their sum (Table \ref{tab:mass}). The open circle adopts for the summed total mass that estimated by \cite{Shaya2017}. The blue band shows range of Milky Way mass estimates \citep{BG2016,Wang2020}. Lines are the stellar mass–halo mass relations from abundance matching \cite{Behroozi2013,Moster2013,Kravtsov2018,Mowla2019}. The yellow box shows the timing mass of the Local Group \citep{vdM2012}. There is a pronounced discrepancy between dynamical and abundance matching estimates of the halo masses of the Local Group and its brightest members.}
    \label{fig:LGmass}
  \end{figure}

Typical AM relations are well calibrated around the knee of the luminosity function, but not to the very low masses of dwarf spheroidals \citep{BBK}. 
The agreement between kinematic mass estimates and AM masses is reasonably good for intermediate mass galaxies like M33.
The two diverge dramatically for the brightest galaxies of the Local Group, M31 and the Milky Way (Table \ref{tab:mass} and Fig.\ \ref{fig:LGmass})

Dynamical studies typically suggest that the total mass of the Milky Way is in the range of 0.5 --- $2 \times 10^{12}\;\Msun$ \citep{BG2016,Wang2020}.
A similar range holds for Andromeda \citep{Chemin2009,Kafle2018}.
AM predicts considerably higher masses: $\sim 3 \times 10^{12}\;\Msun$ for the Milky Way and $> 10^{13}\;\Msun$ for M31. There is some room to hedge this
for the Milky Way by adopting a stellar mass on the low end of the plausible range, but even then the halo mass is abnormally low. The result for Andromeda
differs by an order of magnitude. There is of course some scatter in the stellar mass--halo mass relation from AM, but it is not sufficient to explain 
this stark discrepancy.

According to AM, a galaxy the mass of M31 should be the central object of a much larger group. Indeed, for its stellar mass, the Milky Way should be the central
object of the Local Group. Andromeda should not be here. Despite its normalcy in other respects, the Local Group has far too many stars for its mass from the
perspective of abundance matching. 

It is not clear what the resolution of this issue might be. There could be a morphological type dependence that is not accounted for in the AM relations considered here. 
There might be more mass at large radii than kinematic tracers have so far revealed, though this is limited by the timing argument \citep{vdM2012,Penarrubia,Shaya2017}. 
The discrepancy appears to be real, and would hardly be the first indication of a problem with the dark matter paradigm in the dynamics of galaxies \citep{SM2002,FM2012,2020Galax,BZ2022}.


\begin{discussion}

\discuss{M\"uller}{What is the reason for these strong regularities?}

\discuss{McGaugh}{This behavior was predicted by MOND \citep{MOND}.}

\end{discussion}

\end{document}